\documentclass[12pt,letterpaper]{article}
\pdfoutput=1
\usepackage{jheppub}
\usepackage{amsfonts, amsthm}
\usepackage[english]{babel}
\usepackage[utf8]{inputenc}
\usepackage{slashed}
\usepackage{mathrsfs}
\usepackage{amssymb}
\usepackage{color}
\hypersetup{unicode}

\newcommand{\eq}{\begin{equation}}
\newcommand{\feq}{\end{equation}}
\newcommand{\eqn}{\begin{eqnarray}}
\newcommand{\feqn}{\end{eqnarray}}

\newcommand{\ma}[1]{\mbox{$\mathcal{#1}$}}
\newcommand{\mas}[1]{\mbox{$\mathscr{#1}$}}
\newcommand{\masf}[1]{\mbox{$\mathsf{#1}$}}
\newcommand{\D}{{\rm d}}

\title{Black string first order flow in $N=2$, $d=5$ abelian gauged supergravity}

\author{Dietmar Klemm,}
\author{Nicol\`o Petri}
\author{and Marco Rabbiosi}

\affiliation{Dipartimento di Fisica, Universit\`a di Milano, and \\
INFN, Sezione di Milano, \\
Via Celoria 16, I-20133 Milano, Italy.}

\emailAdd{dietmar.klemm@mi.infn.it}
\emailAdd{nicolo.petri@mi.infn.it}
\emailAdd{marco.rabbiosi@mi.infn.it}
\preprint{IFUM-1052-FT}

\abstract{We derive both BPS and non-BPS first-order flow equations for magnetically charged black
strings in five-dimensional $N=2$ abelian gauged supergravity, using the Hamilton-Jacobi formalism.
This is first done for the coupling to vector multiplets only and $\text{U}(1)$ Fayet-Iliopoulos (FI) gauging,
and then generalized to the case where also hypermultiplets are present, and abelian symmetries of the
quaternionic hyperscalar target space are gauged. We then use these results to derive the attractor
equations for near-horizon geometries of extremal black strings, and solve them explicitely for
the case where the constants appearing in the Chern-Simons term of the supergravity action
satisfy an adjoint identity.
This allows to compute in generality the central charge of the two-dimensional conformal field theory that
describes the black strings in the infrared, in terms of the magnetic charges, the CY intersection numbers
and the FI constants. Finally, we extend the $r$-map to gauged supergravity and use it to relate our
flow equations to those in four dimensions.}


\begin{document}
\maketitle
\flushbottom

\section{Introduction}

Exact solutions to supergravity theories, like black holes, domain walls, plane-fronted waves
etc.~have been instrumental in various developments of string theory, for instance in holography
or black hole microstate counting. Generically, one is interested both in supersymmetric backgrounds
and in solutions that break supersymmetry, like nonextremal black holes. The latter play an important
role for example in holographic descriptions of condensed matter systems at finite temperature, or
of the quark-gluon plasma.

The supergravity equations of motion are coupled, nonlinear, second-order partial differential
equations, and as such quite difficult to solve analytically, even in presence of a high degree of symmetry.
A possible way out is to consider instead the Killing spinor equations, which are of first order in
derivatives, and imply (at least when the Killing vector constucted as a bilinear from the Killing spinor
is timelike) the second-order equations of motion. In this way, however, one obtains only
supersymmetric solutions, and therefore interesting objects like nonextremal or extremal non-BPS black
holes are excluded a priori.

A more general possibility that still involves solving first-order equations, is the Hamilton-Jacobi
approach. This includes the (symmetry-reduced) Killing spinor equations as a special subcase, but is
quite easily generalizable to extremal non-BPS- or even nonextremal black holes. Using Hamilton-Jacobi
theory is essentially\footnote{`Essentially' means that in many flow equations obtained in the literature
by squaring an action, the rhs of \eqref{flow-intro} is not a gradient, or, in other words, the flow is
not driven by a (fake) superpotential (no gradient flow).}
equivalent to writing the action as a sum of squares, which can be seen as follows:
Suppose that, owing to various symmetries (like e.g.~staticity and spherical symmetry), the supergravity
action can be dimensionally reduced to just one dimension\footnote{When there is less symmetry, e.g.~for
rotating black holes, one obtains a field theory living in two or more dimensions, instead of a mechanical
system \cite{Andrianopoli:2012ee,Hristov:2012nu}. In this case, the Hamilton-Jacobi formalism has to be
generalized to the so-called De Donder-Weyl-Hamilton-Jacobi theory \cite{Andrianopoli:2012ee}.}, such that
\begin{equation}
I = \int\D r\left[\frac12{\mathscr G}_{\Lambda\Sigma}\dot q^\Lambda\dot q^\Sigma - U(q)\right]\,,
\label{action-intro}
\end{equation}
where $r$ is a radial variable (the `flow' direction), the $q^\Lambda(r)$ denote collectively the dynamical
variables, $U(q)$ is the potential and ${\mathscr G}_{\Lambda\Sigma}(q)$ the metric on the
target space parametrized by the $q^\Lambda$, with inverse ${\mathscr G}^{\Lambda\Sigma}$.
Now suppose that $U$ can be expressed in terms of a (fake) superpotential $W$ as
\begin{equation}
U = E - \frac12{\mathscr G}^{\Lambda\Sigma}\frac{\partial W}{\partial q^\Lambda}
\frac{\partial W}{\partial q^\Sigma}\,, \label{Ham-Jac-intro}
\end{equation}
where $E$ is a constant. Then, the action \eqref{action-intro} becomes
\begin{equation}
I = \int\D r\left[\frac12{\mathscr G}_{\Lambda\Sigma}\left(\dot q^\Lambda - {\mathscr G}^{\Lambda
\Omega}\frac{\partial W}{\partial q^\Omega}\right)\left(\dot q^\Sigma - {\mathscr G}^{\Sigma
\Delta}\frac{\partial W}{\partial q^\Delta}\right) + \frac{\D}{\D r}(W - Er)\right]\,,
\end{equation}
which is up to a total derivative equal to
\begin{equation}
I = \int\D r\,\frac12{\mathscr G}_{\Lambda\Sigma}\left(\dot q^\Lambda - {\mathscr G}^{\Lambda
\Omega}\frac{\partial W}{\partial q^\Omega}\right)\left(\dot q^\Sigma - {\mathscr G}^{\Sigma
\Delta}\frac{\partial W}{\partial q^\Delta}\right)\,.
\end{equation}
The latter is obviously stationary if the first-order flow equations
\begin{equation}
\dot q^\Lambda = {\mathscr G}^{\Lambda\Omega}\frac{\partial W}{\partial q^\Omega} \label{flow-intro}
\end{equation}
hold. But \eqref{Ham-Jac-intro} is nothing else than the reduced Hamilton-Jacobi equation, with
$W$ Hamilton's characteristic function, while \eqref{flow-intro} represents the expression for the
conjugate momenta
$p_\Lambda=\partial{\mathscr L}/\partial\dot q^\Lambda={\mathscr G}_{\Lambda\Sigma}\dot q^\Sigma$
in Hamilton-Jacobi theory\footnote{For further discussions of the relationship between the Hamilton-Jacobi
formalism and the first-order equations derived from a (fake) superpotential 
cf.~\cite{Andrianopoli:2009je,Trigiante:2012eb}.}.

First-order flow equations, derived either by writing the dimensionally reduced action as
a sum of squares or from the Hamilton-Jacobi formalism, appear for many different settings in the
literature, both in ungauged and gauged supergravity, and for BPS-, extremal non-BPS- and even 
nonextremal black holes, cf.~e.g.~\cite{Miller:2006ay,Ceresole:2007wx,LopesCardoso:2007qid,
Cardoso:2008gm,Andrianopoli:2009je,Dall'Agata:2010gj,Galli:2011fq,Barisch:2011ui,Trigiante:2012eb,
BarischDick:2012gj,Klemm:2012vm,Gnecchi:2012kb,Gnecchi:2014cqa,Cardoso:2015wcf,Klemm:2016wng}
for an (incomplete) list of references. In particular, ref.~\cite{Ceresole:2001wi} establishes general
properties of supersymmetric flow equations for domain walls in five-dimensional $N=2$ gauged
supergravity coupled to vector- and hypermultiplets.

Here we shall consider magnetically charged black strings in five-dimensional $N=2$ gauged supergravity,
and obtain first-order flow equations for them. This is done first for the Fayet-Iliopoulos (FI)-gauged
case, and then generalized to include also hypermultiplets, where abelian symmetries of the
quaternionic hyperscalar target manifold are gauged. Extremal magnetic black strings interpolate
between (so-called `magnetic') AdS$_5$ at infinity and $\text{AdS}_3\times\Sigma$ (with $\Sigma$
a two-dimensional space of constant curvature) at the horizon. Holographically, this corresponds
to an RG flow across dimensions from a 4d field theory to a two-dimensional CFT in the
infrared \cite{Maldacena:2000mw,Benini:2012cz,Benini:2015bwz,Karndumri:2013iqa,Amariti:2016mnz}.
By plugging the near-horizon data into the flow equations, one
gets the attractor equations. We solve the latter in full generality under the additional assumption
that the `adjoint identity' \eqref{adjoint-id} holds. This enables us to compute the central charge
of the 2d CFT that describes the black strings in the IR, in terms of the string charges, the FI parameters
(or, more generally, the moment maps, if also hypermultiplets are present),
and the constants $C_{IJK}$ that appear in the Chern-Simons term of the supergravity action.

The remainder of this paper is organized as follows: In the next section, we briefly review $N=2$,
$d=5$ Fayet-Iliopoulos gauged supergravity. In section \ref{BPSflow}, we derive the BPS flow
equations for static black strings, and generalize them in \ref{nonBPSflow} to the non-supersymmetric
case, by using a simple deformation of the BPS superpotential. After that, in section \ref{hyper},
the presence of (abelian) gauged hypermultiplets is taken into account as well. In \ref{entropy}, the
flow equations are solved in the near-horizon limit, where the geometry contains an $\text{AdS}_3$
factor. This leads to the attractor equations for black strings, that we subsequently solve in full
generality, and compute the central charge of the 2d CFT that describes the black strings in the IR.
We conclude in section \ref{sec:final} with some proposals for extensions of our work.
An appendix contains some useful relations in very special geometry and the construction of the $r$-map
in the gauged case.

\newpage


\section{\label{Setup}$N=2$, $d=5$ Fayet-Iliopoulos gauged supergravity}

The bosonic Lagrangian of $N=2$, $d=5$ FI-gauged supergravity coupled to $n_{\text v}$ vector
multiplets is given by \cite{Gunaydin:1983bi,Gunaydin:1984ak}\footnote{The indices $I,J,\ldots$ range
from $1$ to $n_{\text v}+1$, while $i,j,\ldots=1,...,n_{\text v}$.}
\begin{equation}
e^{-1}\mathscr L = \frac12 R - \frac12\mathcal G_{ij}\partial_{\mu}\phi^i\partial^{\mu}\phi^j - \frac14
G_{IJ} F^I_{\mu\nu} F^{J\mu\nu} + \frac{e^{-1}}{48} C_{IJK}\epsilon^{\mu\nu\rho\sigma\lambda}
F^I_{\mu\nu} F^J_{\rho\sigma} A^K_{\lambda} - \mathrm{g}^2 V\,,
\label{eq:genlag}
\end{equation}
where the scalar potential reads
\begin{equation}
V = V_I V_J\left(\frac92\mathcal G^{ij}\partial_i h^I\partial_j h^J - 6 h^I h^J\right)\,.
\end{equation}
Here, $V_I$ are FI constants, $\partial_i$ denotes a partial derivative with respect to the real scalar field
$\phi^i$, and $h^I=h^I(\phi^i)$ satisfy the condition
\begin{equation}
\mathcal V\equiv\frac16 C_{IJK} h^I h^J h^K= 1\,.
\label{eq:intersection}
\end{equation}
Moreover, $G_{IJ}$ and $\mathcal G_{ij}$ can be expressed in terms of the homogeneous cubic polynomial
$\mathcal V$ which defines a `very special geometry' \cite{deWit:1992cr},
\begin{equation}
G_{IJ} = -\frac12\frac{\partial}{\partial h^I}\frac{\partial}{\partial h^J}\left.\log\mathcal V\right|_{\mathcal
V=1}\,, \qquad \mathcal G_{ij} = \partial_i h^I\partial_j h^J\left.G_{IJ}\right|_{\mathcal V=1}\,.
\end{equation}
Further useful relations can be found in appendix \ref{app-very-special}. We note that if the
five-dimensional theory is obtained by gauging a supergravity theory coming from a Calabi-Yau
compactification of M-theory, then $\mathcal V$ is the intersection form, $h^I$ and
$h_I\equiv\frac16 C_{IJK}h^J h^K$ correspond to the size of the two- and four-cycles and the constants
$C_{IJK}$ are the intersection numbers of the Calabi-Yau threefold \cite{Cadavid:1995bk}.

\section{\label{BPSflow}BPS flow for a static black string}

Very special real K\"ahler manifolds can be viewed as the pre-image of the supergravity
$r$-map \cite{Berkooz:2008rj,Cortes:2011aj}. This suggests to consider the five-dimensional
spacetime as a Kaluza-Klein uplift of the usual static black holes in four dimensions. Moreover, a pure
string solution in $d=5$ supports only magnetic charges, thus the field configuration reads
\begin{equation}
\begin{split}
&\D s^2 = e^{2T(r)}\D z^2 + e^{-T(r)}\left(-e^{2U(r)}\D t^2 + e^{-2U(r)}\D r^2 + e^{2\psi(r) - 2U(r)}
\D\sigma^2_\kappa\right)\,, \\
& F^I = p^I f_\kappa (\theta)\D\theta\wedge\D\phi\,, \qquad\phi^i = \phi^i(r)\,, \label{eq:ans}
\end{split}
\end{equation}
where $\D\sigma_{\kappa}^2=\D\theta^2+f_{\kappa}^2(\theta)\D\varphi^2$ is the metric on the
two-dimensional surfaces $\Sigma=\{\mathrm{S}^2,\mathrm{H}^2\}$ of constant scalar
curvature $R=2\kappa$, with $\kappa\in\{1,-1\}$, and
\begin{equation}
f_\kappa(\theta) = \frac{1}{\sqrt{\kappa}} \sin(\sqrt{\kappa}\theta) = 
\left\{\begin{array}{c@{\quad}l} \sin\theta\, & \kappa=1\,, \\
                                                 \sinh\theta\, & \kappa=-1\,. \end{array}\right.
\end{equation}
Plugging the ansatz \eqref{eq:ans} into the equations of motion following from \eqref{eq:genlag} yields
a set of ordinary differential equations that can be derived from the one-dimensional effective action
\begin{equation}
\begin{split}
&S_{\text{eff}} = \int\D r\left[e^{2\psi}\left(U^{\prime 2} + \frac34 T^{\prime 2} - \psi^{\prime 2} +
\frac12\mathcal G_{ij}\phi^{\prime\,i}\phi^{\prime\,j}\right) - V_{\text{eff}}\right]\,, \\
&V_{\text{eff}} = \kappa - e^{2\psi - 2U - T}\mathrm{g}^2 V - \frac12 e^{2U + T - 2\psi}G_{IJ} p^I p^J\,,
\end{split}
\end{equation}
imposing in addition the zero energy condition $H_{\text{eff}} = 0$. In the Hamilton-Jacobi formulation
the latter becomes the partial differential equation
\begin{equation}
e^{-2\psi}\left((\partial_U W)^2 - (\partial_\psi W)^2 + \frac 43(\partial_T W)^2 + 2 \mathcal G^{ij}
\partial_i W\partial_j W\right) + V_{\text{eff}} = 0\,, \label{eq:hjbs}
\end{equation} 
where $W$ is Hamilton's characteristic function that we will sometimes refer to also as (fake)
superpotential.

A solution of (\ref{eq:hjbs}) permits to write the action as a sum of squares and to derive a set of
first-order flow equations by setting the squares to zero\footnote{As explained in the introduction,
these are of course equivalent to the
usual first-order equations in the Hamilton-Jacobi formalism.}. Guided by the four-dimensional
case \cite{Dall'Agata:2010gj,Klemm:2012vm,Klemm:2016wng},
the ansatz for the simplest non-trivial solution is
\begin{equation}
W = a e^{U + \frac T2} p^I h_I + b e^{2\psi - U - \frac T2} V_I h^I\,, \label{eq:mainHJ}
\end{equation}
where $a,b$ are constants. Using (\ref{eq:veryspecial}), one can show that (\ref{eq:mainHJ}) solves (\ref{eq:hjbs}) if one imposes 
\begin{equation}
a= -\frac 34\,, \qquad b = \frac32\mathrm{g}\,, \qquad V_I p^I = -\frac{\kappa}{3\mathrm{g}}\,.
\label{eq:HJcoefficients}
\end{equation}
The last of \eqref{eq:HJcoefficients} is a sort of Dirac quantization condition for the linear combination
$V_Ip^I$ of the magnetic charges in terms of the inverse gauge coupling constant $\mathrm{g}^{-1}$.
This solution for $W$ leads to the first-order flow 
\begin{equation}
\begin{split}
& U^{\prime} = -\frac34 e^{U + \frac T2 - 2\psi} p^I h_I - \frac32\mathrm{g}\,e^{-U -\frac T2} V_I h^I\,,\\
&\psi^{\prime} = -3\mathrm{g}\, e^{-U -\frac T2} V_I h^I\,, \qquad T^{\prime} = \frac23 U^{\prime}\,, \\
&\phi^{\prime\,i} = 3\mathcal G^{ij}\left(-\frac12 e^{U + \frac T2 - 2\psi} p^I\partial_j h_I +
\mathrm{g}\,e^{-U -\frac T2} V_I\partial_j h^I\right)\,. \label{eq:flow}
\end{split}
\end{equation}
One can check that \eqref{eq:flow} coincides with the system obtained in \cite{Cacciatori:2003kv} from
the Killing spinor equations. In particular, it is easy to verify that the supersymmetric magnetic black
string solution of \cite{Klemm:2000nj} satisfies \eqref{eq:flow}. Moreover, introducing a new radial coordinate $R$ and the warp factors $f$ and $\rho$ such that
\begin{equation}
U = \frac32 f\,, \qquad \psi = 2f + \rho\,, \qquad T = f\,, \qquad \frac{\D R}{\D r} = e^{-3f}\,,
\label{eq:maldaansatz}
\end{equation}
and specifying to the stu model, one shows that \eqref{eq:flow} is precisely the system of equations derived
in appendix 7.1 of~\cite{Maldacena:2000mw} from the Killing spinor equations.

\section{Non-BPS flow}
\label{nonBPSflow}

One of the main advantages of the Hamilton-Jacobi formalism is to allow for a simple generalization 
of the first-order flow driven by \eqref{eq:mainHJ} to a non-BPS one. Similar to the case of $N=2$,
$d=4$ abelian gauged supergravity \cite{Gnecchi:2012kb, Klemm:2016wng}, one introduces
a `field rotation matrix' ${S^I}_J$ such that
\begin{equation}
G_{LK} {S^L}_I {S^K}_J = G_{IJ}\,. \label{eq:rotmat}
\end{equation}
A nontrivial $S$ (different from $\pm\text{Id}$) allows to generate new solutions from known ones by
‘rotating charges’. This technique was first introduced in \cite{Ceresole:2007wx,LopesCardoso:2007qid},
and generalizes the sign-flipping procedure of \cite{Ortin:1996bz}.
Using \eqref{eq:rotmat}, one easily verifies that
\begin{equation}
\tilde W = -\frac43 e^{U + \frac T2}  h_I {S^I}_J p^J + \frac23\mathrm{g} e^{2\psi - U - \frac T2} V_I h^I
\end{equation}
satisfies again the Hamilton-Jacobi equation \eqref{eq:hjbs}, provided the modified quantization condition
\begin{equation}
V_I {S^I}_J p^J = -\frac{\kappa}{3\mathrm{g}}
\end{equation}
holds. This leads to the first-order flow driven by $\tilde W$,
\begin{equation}
\begin{split}
& U^{\prime} = -\frac34 e^{U + \frac T2 - 2\psi} h_I {S^I}_J p^J - \frac32\mathrm{g}\, e^{-U - \frac T2}
V_I h^I\,,\\
& \psi^{\prime} = -3\mathrm{g}\, e^{-U - \frac T2} V_I h^I\,, \qquad T^{\prime} = \frac23 U^{\prime}\,, \\
& \phi^{\prime\,i} = 3\mathcal G^{ij}\left(-\frac12 e^{U + \frac T2 - 2\psi}\partial_j h_I  {S^I}_J p^J + \mathrm{g}\, e^{-U - \frac T2} V_I \partial_j h^I\right)\,.
\end{split}
\end{equation} 
An interesting example for which \eqref{eq:rotmat} admits nontrivial solutions, is the model
$\mathcal V = h^1 h^2 h^3 = 1$ (cf.~e.g.~\cite{Cacciatori:2003kv}), for which
\begin{equation}
G_{IJ} = \frac{\delta_{IJ}}{2(h^{I})^2}\,.
\end{equation}
In this case a particular solution of \eqref{eq:rotmat} is given by
\begin{equation}
{S^I}_J =
\begin{pmatrix}
\epsilon_1   &  0  &  0 \\
0  &  \epsilon_2 &  0 \\
0  &  0  &  \epsilon_3
\end{pmatrix}\,,
\end{equation}
with $\epsilon_I=\pm 1$. These matrices form a discrete subgroup
$D=(\mathbb Z_2)^3\subset\text{GL}(3,\mathbb R)$. Since there are two equivalent BPS branches,
the independent solutions correspond to elements of the quotient group $D/\mathbb Z_2$.

\section{Inclusion of hypermultiplets}
\label{hyper}

We now generalize our analysis to include also the coupling to $n_{\text H}$ hypermultiplets. The
charged hyperscalars $q^u$ ($u=1,\cdots,4n_{\text H}$) parametrize a quaternionic K\"ahler manifold
with metric $h_{uv}(q)$, i.e., a $4n_{\text H}$-dimensional Riemannian manifold admitting a locally
defined triplet $\vec{K}_u^{\phantom{u}v}$ of almost complex structures satisfying the quaternion relation
\begin{equation}
\label{eq:quaternionic_kahler_complexstruct_definition}
h^{st} K^x_{\phantom{x}us} K^y_{\phantom{y}tw} = -\delta^{xy} h_{uw} + \varepsilon^{xyz}
K^z_{\phantom{z}uw}\,,
\end{equation}
and whose Levi-Civita connection preserves $\vec{K}$ up to a rotation,
\begin{equation}
\label{eq:quaternionic_kahler_complexstruct_rotation}
\nabla_w {\vec K}_u^{\phantom{u}v} + \vec{\omega}_w\times{\vec K}_u^{\phantom{u}v} = 0\,,
\end{equation}
where $\vec\omega\equiv \vec\omega_u (q)\, dq^u$ is the connection of the $\mathrm{SU}(2)$-bundle for which the quaternionic manifold is the base. 
The $\mathrm{SU}(2)$ curvature is proportional to the complex structures,
\begin{equation}
\label{eq:quaternionkahl_su2curv}
\Omega^x\equiv\D\omega^x + \frac12\varepsilon^{xyz}\omega^y\wedge\omega^z = -K^x\,.
\end{equation} 
Here we shall consider only gaugings of abelian isometries of the quaternionic K\"ahler metric $h_{uv}$.
These are generated by commuting Killing vectors $k_I^u (q)$. For each Killing vector one can introduce
a triplet of moment maps, $P_I^x$, such that
\begin{equation}
\label{eq:mommaps}
\masf{D}_u P_I^x\equiv \partial_u P_I^x + \varepsilon^{xyz}\omega_{\phantom{y} u}^y
P_I^z = -2\Omega^x{}_{uv} k_I^v\,.
\end{equation}
One of the most important relations satisfied by the moment maps is the so-called equivariance relation.
For abelian gaugings it has the form
\begin{equation}
\frac12\epsilon^{xyz} P^y_I P^z_J - \Omega^x_{uv} k^u_I k^v_J =  0\,.
\label{eq:equivariance}
\end{equation}
The bosonic Lagrangian is now given by\footnote{\eqref{eq:genlaghyper} can be obtained from the
Lagrangian in \cite{Ceresole:2001wi} by rescaling $a_{IJ}\to\frac23 G_{IJ}$, $C_{IJK}\to\frac16 C_{IJK}$,
$k_I\to 2k_I$, $A^I\to\sqrt{\frac32}A^I$, $g\to\sqrt{\frac32}g$.}
\begin{eqnarray}
e^{-1}\mathscr L &=&\frac12 R - \frac12\mathcal G_{ij}\partial_{\mu}\phi^i\partial^{\mu}\phi^j - 
h _{uv}\hat{\partial}_{\mu}q^u\hat{\partial}^{\mu} q^v - \frac14 G_{IJ} F^I_{\mu\nu} F^{J\mu\nu}
\nonumber \\
&&+ \frac {e^{-1}}{48} C_{IJK}\epsilon^{\mu\nu\rho\sigma\lambda} F^I_{\mu\nu} F^J_{\rho\sigma}
A^K_{\lambda} - \mathrm{g}^2 V\,, \label{eq:genlaghyper}
\end{eqnarray}
with the covariant derivative
\begin{equation}
\hat{\partial}_{\mu}q^u = \partial_{\mu}q^u + 3\mathrm{g} A_{\mu}^I k_I^u\,,
\end{equation}
and the scalar potential
\begin{equation}
V = P_I^x P_J^x\left(\frac92\mathcal G^{ij}\partial_i h^I\partial_j h^J - 6 h^I h^J\right) + 9 h_{uv} k_I^u
k_J^v h^I h^J\,. \label{eq:pothyper}
\end{equation}
Varying \eqref{eq:genlaghyper} w.r.t~$A_{\mu}^I$, one obtains the Maxwell equations
\begin{equation}
\partial_{\mu}\left(e\,\ma G_{IJ} F^{J\mu\nu}\right) + \frac14\epsilon^{\mu\lambda\rho\sigma\nu}C_{IJK}
\partial_{\mu} A_{\lambda}^J\partial_{\rho} A_{\sigma}^K = 6\mathrm{g} e h_{uv} k_I^u
\hat{\partial}^{\nu}q^v\,. \label{eq:max}
\end{equation}
Imposing the ansatz \eqref{eq:ans}, the $t$-, $\theta$- and $z$-components of \eqref{eq:max} are automatically satisfied, while the $r$- and $\varphi$-components become respectively
\begin{equation}
h_{uv} k_I^u q^{\prime\,v} = 0\,, \qquad k_I^u p^I = 0\,.
\end{equation}
The remaining equations of motion can be derived from the effective action
\begin{equation}
\begin{split}
&S_{\text{eff}} = \int\D r\left[e^{2\psi}\left(U^{\prime 2} + \frac34 T^{\prime 2} - \psi^{\prime 2} +
\frac12\mathcal G_{ij}\phi^{\prime\,i}\phi^{\prime\,j} +  h_{uv} q^{\prime\,u} q^{\prime\,v}\right) -
V_{\text{eff}}\right]\,, \\
&V_{\text{eff}} = \kappa - e^{2\psi - 2U - T}\mathrm{g}^2 V - \frac12 e^{2U + T - 2\psi}G_{IJ} p^I p^J\,,
\end{split}
\end{equation}
supplemented by the Hamiltonian constraint $H_{\text{eff}}=0$. The latter leads to the
Hamilton-Jacobi equation
\begin{equation}
e^{-2\psi}\left((\partial_U W)^2 - (\partial_\psi W)^2 + \frac43(\partial_T W)^2 + 2\mathcal G^{ij}\partial_i
W\partial_j W +  h^{uv}\partial_u W\partial_v W\right) + V_{\text{eff}} = 0\,. \label{eq:hjbshyper}
\end{equation}
Guided by the FI-gauged case and by previous work in four dimensions \cite{Klemm:2016wng}, we use the
ansatz
\begin{equation}
W = c e^{U + \frac T2}\ma Z + d e^{2\psi - U - \frac T2}\ma L\,, \label{eq:mainHJhyper}
\end{equation}
where
\begin{equation}
\ma Z = p^I h_I\,, \qquad \ma L = \ma Q^x\ma W^x\,, \qquad \ma Q^x = p^I P^x_I\,, \qquad
\ma W^x = h^J P_J^x\,. \label{def-ZLQxWx}
\end{equation}
Using some relations of very special geometry as well as
\eqref{eq:quaternionic_kahler_complexstruct_definition}, \eqref{eq:quaternionkahl_su2curv} and
\eqref{eq:equivariance}, one can show that \eqref{eq:mainHJhyper} solves indeed \eqref{eq:hjbshyper}
provided that
\begin{equation}
c = -\frac34\,, \qquad d = -\frac92\kappa\mathrm{g}^2\,, \qquad \ma Q^x\ma Q^x =
\frac1{9\mathrm{g}^2}\,.
\end{equation}
The solution \eqref{eq:mainHJhyper} leads then to the first-order flow equations
\begin{equation}
\begin{split}
& U^{\prime} = -\frac34 e^{U + \frac T2 - 2\psi}\ma Z + \frac92\kappa\mathrm{g}^2 e^{-U -\frac T2}
\ma L\,, \qquad T^{\prime} = \frac23 U^{\prime}\,, \\
&\psi^{\prime} = 9\kappa\mathrm{g}^2 e^{-U - \frac T2}\ma L\,, \\
&\phi^{\prime\,i} = \mathcal G^{ij}\left(-\frac32 e^{U + \frac T2 - 2\psi}\partial_j\ma Z - 9\kappa
\mathrm{g}^2 e^{-U - \frac T2}\partial_j\ma L\right)\,, \\
& q^{\prime\,u} = -\frac 92 \kappa\mathrm{g}^2 e^{-U - \frac T2} h^{uv}\partial_v\ma L\,.
\label{eq:hyperflow}
\end{split}
\end{equation}
One can recast \eqref{eq:hyperflow} into a form very similar to that of the first-order flow in four
dimensions, cf.~eqns.~(3.43) in \cite{Klemm:2016wng}. Integrating $T^{\prime} = \frac23 U^{\prime}$
and plugging this into the remaining equations of  \eqref{eq:hyperflow}, one gets
\begin{equation}
\begin{split}
& T^{\prime} = -\frac12 e^{2T - 2\psi}\ma Z + 3\kappa\mathrm{g}^2 e^{- 2T}\ma L\,, \\
&\psi^{\prime} = 9\kappa\mathrm{g}^2 e^{-2T}\ma L\,, \\
&\phi^{\prime\,i} = \mathcal G^{ij}\left(-\frac32 e^{2T - 2\psi}\partial_j\ma Z - 9\kappa\mathrm{g}^2
e^{-2T}\partial_j\ma L\right)\,, \\
& q^{\prime\,u} = -\frac 92 \kappa\mathrm{g}^2 e^{-2T} h^{uv}\partial_v\ma L\,. \label{eq:hyperflow1}
\end{split}
\end{equation}
Using the equation for $\phi^{\prime\,i}$ together with $(h^I)^\prime=\phi^{\prime\,i}\partial_i h^I$ and
\eqref{eq:veryspecial1}, the equations for $T$ and $\phi^i$ can be rewritten as
\begin{equation}
e^{2\psi}\left(e^{-2T} h^I\right)^\prime + 9\mathrm{g}^2\kappa e^{2\psi-4T}\ma Q^x P^x_J G^{IJ}-p^I
= 0\,. \label{eq:hyperflow2}
\end{equation}
Note that the FI case can be recovered imposing $P^1_I=P^2_I=0$ and $P^3_I=V_I$. Then the charge
quantization condition $\ma Q^x\ma Q^x=1/(9\mathrm{g}^2)$ boils down to
$\ma Q^3=p^IV_I=\pm\kappa/(3\mathrm{g})$ (use $\kappa^2=1$),
while $\ma L$ in \eqref{def-ZLQxWx} becomes
$\ma L=\pm\frac{\kappa}{3\mathrm{g}} h^JV_J$, which is the expression appearing in \eqref{eq:mainHJ}.
The two signs correspond to the two equivalent
BPS branches; in section \ref{BPSflow} the lower sign was chosen.

\section{Attractors and central charge of the dual CFT}
\label{entropy}

In this section we want to investigate the near-horizon configurations of the black string. To keep
things simple, we shall first concentrate on the hyperless FI-gauged case considered in section \ref{BPSflow},
and set $\mathrm{g}=1$. The geometry is of the type $\mathrm{AdS}_3\times\Sigma$ with $\Sigma=\{\text{S}^2,\text{H}^2\}$, and we assume that the scalars stabilize regularly at the horizon, i.e.,
$\phi^{\prime\,i}=0$. Note that a similar problem was solved in four dimensions in \cite{Halmagyi:2013qoa}
for the case of symmetric special K\"ahler manifolds with cubic prepotential\footnote{Supersymmetric
Bianchi attractors in $N=2$, $d=5$ gauged supergravity coupled to vector- and hypermultiplets were
analyzed recently in \cite{Chakrabarty:2016mpf}.}.

In the coordinates $(t,R,z,\theta,\phi)$, where $R$ was introduced in \eqref{eq:maldaansatz},
the metric \eqref{eq:ans} takes the form
\begin{equation}
\D s^2 = e^{2f}(-\D t^2 + \D R^2 + \D z^2) + e^{2\rho}\D\sigma^2_{\kappa}\,,
\end{equation}
and the first-order flow equations \eqref{eq:flow} become
\begin{equation}
\begin{split}
& f^\prime = -e^f(h^I V_I + \frac12 e^{-2\rho}\ma Z )\,, \\
&\rho^\prime = -e^f(h^I V_I - e^{-2\rho}\ma Z)\,, \\
&\phi^{\prime\,i} = 3\mathcal G^{ij} e^f (\partial_j h^I V_I - \frac12 e^{-2\rho}\partial_j\ma Z)\,,
\label{eq:flowequation2}
\end{split}
\end{equation}
where the primes now denote derivatives w.r.t.~$R$.
For a product space $\mathrm{AdS}_3\times\Sigma$ we have
\begin{equation}
e^{2f} = \frac{R_{\mathrm{AdS}_3}^2}{R^2}\,, \qquad e^{2\rho} = R_{\text H}^2\,. \label{eq:nhans}
\end{equation}
Plugging this together with $\phi^{\prime\,i}=0$ into \eqref{eq:flowequation2}, one obtains a system of
algebraic equations whose solution fixes the near-horizon values of the scalars in terms of the charges and
the FI parameters,
\begin{equation}
h^I V_I = \frac2{3R_{\mathrm{AdS}_3}}\,, \qquad \ma Z = R_{\text H}^2 h^I V_I\,, \qquad
\partial_i\ma Z = 2R_{\text H}^2\partial_i h^I V_I\,. \label{eq:attractor}
\end{equation}
For the ansatz \eqref{eq:nhans}, the FI-version of \eqref{eq:hyperflow2} (obtained by taking
$\ma Q^xP^x_J=\ma Q^3P^3_J=-\kappa V_J/(3\mathrm{g})$) reduces to
\begin{equation}
e^{f + 2\rho}(e^{-2f} h^I)^\prime - 3 e^{2\rho} G^{IJ}V_J - p^I = 0\,.
\end{equation}
Using \eqref{eq:nhans} and \eqref{eq:attractor}, this can be rewritten as
\begin{equation}
p^I + 3 R_{\text H}^2 G^{IJ} V_J = 3\ma Z h^I\,. \label{eq:attractor2}
\end{equation}
We want to solve the attractor equations \eqref{eq:attractor} (or equivalently \eqref{eq:attractor2}) in
order to express $R_{\mathrm{AdS}_3}$, $R_{\text H}$ and $h^I$ in terms of $p^I$ and $V_I$. To this end,
contract the third relation of \eqref{eq:veryspecial1} with $V_I$ to get
\begin{equation}
\mathcal G^{ij}\partial_i h^I V_I\partial_j h_J = -\frac23 V_J + \frac23 h^I V_I h_J\,. \label{eq:veryspecialL}
\end{equation}
With \eqref{eq:attractor}, this becomes
\begin{equation}
R_{\text H}^2 V_J = -\frac34\mathcal G^{ij}\partial_i\ma Z\partial_j h_J + \ma Z h_J\,.
\end{equation}
Using $h_I=\frac16 C_{IJK}h^Jh^K$ and \eqref{eq:veryspecial1}, one obtains
\begin{equation}
R_{\text H}^2 V_J = \frac16 C_{JKL} p^K h^L\,. \label{RH^2V_J}
\end{equation}
Let us introduce the charge-dependent matrix
\begin{equation}
C_{p\,IJ}\equiv C_{IJK}p^K\,.
\end{equation}
Using the adjoint identity \eqref{adjoint-id}, one easily shows that $C_{p\,IJ}$ is invertible, with inverse
\begin{equation}
C^{IJ}_p = 3\frac{C^{IJK} C_{KMN} p^M p^N - p^I p^J}{C_p}\,, \label{eq:inverseC}
\end{equation}
where $C_p=C_{IJK}p^Ip^Jp^K$. \eqref{RH^2V_J} implies then
\begin{equation}
h^I = 6 R_{\text H}^2 C^{IJ}_p V_J\,. \label{eq:nhrelation}
\end{equation}
Plugging \eqref{eq:nhrelation} into \eqref{eq:intersection}, one can derive a general expression for
$R_{\text H}$ in terms of the intersection numbers, the charges and the FI parameters,
\begin{equation}
R_{\text H}^2 = (36 C_{IJK} C_p^{IM} C_p^{JN} C_p^{KP} V_M V_N V_P)^{-\frac13}\,. \label{eq:entropy}
\end{equation}
Using this in \eqref{eq:nhrelation} gives the values of the scalars at the horizon,
\begin{equation}
h^I = \frac{6 C^{IJ}_p V_J}{(36 C_{KLM} C_p^{KN} C_p^{LP} C_p^{MR} V_N V_P V_R)^{\frac13}}\,.
\label{eq:scalarsnh}
\end{equation}
Contracting \eqref{eq:nhrelation} with $V_I$ and using the first equation of \eqref{eq:attractor} as well as
\eqref{eq:entropy}, we obtain an expression for the $\text{AdS}_3$ curvature radius $R_{\mathrm{AdS}_3}$,
\begin{equation}
R_{\mathrm{AdS}_3} = \frac{(36 C_{IJK} C_p^{IM} C_p^{JN} C_p^{KP} V_M V_N V_P)^{\frac13}}{9C^{RS}_p
V_R V_S}\,. \label{eq:Rads}
\end{equation}
Finally, one can plug \eqref{eq:inverseC} into \eqref{eq:entropy}, \eqref{eq:scalarsnh} and \eqref{eq:Rads},
and use \eqref{adjoint-id} to write the solutions of \eqref{eq:attractor} and \eqref{eq:attractor2} as
\begin{equation}
\begin{split}
& R_{\text H}^2 = (\mas C^{IJK}(p) V_I V_J V_K)^{-\frac13}\,, \\
& h^I = \frac{6\kappa}{C_p}\frac{p^I + 3\kappa C^{IJK} C_{KLM} p^L p^M V_J}{(\mas C^{NPR}(p) V_N V_P
V_R)^{\frac13}}\,, \\
& R_{\mathrm{AdS}_3} = \frac{C_p}{27}\frac{(\mas C^{IJK}(p) V_I V_J V_K)^{\frac13}}
{C^{LMN} C_{NRS}\, p^R p^S V_L V_M - \frac19}\,, \label{eq:solutions}
\end{split}
\end{equation}
where
\begin{equation}
\mas C^{IJK}(p) = -\frac{108}{C_p}\left[2 C^{IJK} - \frac{9}{C_p} p^{(I} C^{JK)M} C_{MNP} p^N p^P +
\frac{9}{C_p} p^I p^J p^K\right]\,.
\end{equation}
The central charge of the two-dimensional conformal field theory that describes the black strings
in the infrared \cite{Maldacena:2000mw,Benini:2013cda,Hristov:2014eza}, is given
by \cite{Brown:1986nw}
\begin{equation}
c = \frac{3R_{\text{AdS}_3}}{2G_3}\,, \label{central}
\end{equation}
where $G_3$ denotes the effective Newton constant in three dimensions, related to $G_5$ by
\begin{equation}
\frac1{G_3} = \frac{R_{\text H}^2\mathrm{vol}(\Sigma)}{G_5}\,. \label{G_3}
\end{equation}
In what follows, we assume $\Sigma$ to be compactified to a Riemann surface of genus $\mathfrak{g}$,
with $\mathfrak{g}=0,2,3,\ldots$. The unit $\Sigma$ has Gaussian curvature $K=\kappa$, and thus
the Gauss-Bonnet theorem gives
\begin{equation}
\mathrm{vol}(\Sigma) = \frac{4\pi(1 - \mathfrak{g})}{\kappa}\,. \label{GB}
\end{equation}
Using \eqref{G_3} and \eqref{GB} in \eqref{central} yields for the central charge
\begin{equation}
c = \frac{6\pi(1 - \mathfrak{g}) R_{\text{AdS}_3} R_{\text H}^2}{\kappa G_5}\,.
\end{equation}
The curvature radii $R_{\text{AdS}_3}$ and $R_{\text H}$ can be expressed in terms of the constants
$C_{IJK}$, the magnetic charges $p^I$ and the FI parameters $V_I$ by means of \eqref{eq:solutions}.
This leads to
\begin{equation}
c = \frac{2\pi(1 - \mathfrak{g}) C_p}{\kappa G_5 (9 C^{IJK} C_{KMN} p^M p^N V_I V_J - 1)}\,. \label{c-final}
\end{equation}
If the hyperscalars are running, one has to consider also the near-horizon limit of the last equation of \eqref{eq:hyperflow1}. Assuming $q^{\prime\,u}=0$ at the horizon and using \eqref{eq:mommaps},
one easily derives the algebraic condition
\begin{equation}
k^u_I h^I=0\,.
\end{equation}
As far as the remaining equations of \eqref{eq:hyperflow1} are concerned, one can follow the same steps
as in this section, with the only difference that $V_I$ has to be replaced everywhere by
$-3\kappa \ma Q^x P_I^x$.

\section{Final remarks}
\label{sec:final}

Let us conclude our paper with the following suggestions for possible extensions and
questions for future work:

\begin{itemize}
\item Try to solve the flow equations in presence of hypermultiplets obtained in section \ref{hyper} for
some specific models, e.g.~like those considered in \cite{Ceresole:2001wi}, to explicitely construct
black strings with running hyperscalars, similar in spirit to the black holes found in \cite{Chimento:2015rra}.
To the best of our knowledge, no such solutions are known up to now.
\item Derive first-order equations for electrically charged black holes (rather than magnetically charged
black strings) in five-dimensional matter-coupled gauged supergravity.
\item Extend our work to the nonextremal case, similar to what was done
in \cite{Miller:2006ay,Cardoso:2008gm,Galli:2011fq,Barisch:2011ui,Gnecchi:2014cqa} in different contexts.
Up to now, the only known nonextremal black string solutions in $\text{AdS}_5$ were constructed
in \cite{Bernamonti:2007bu} for minimal gauged supergravity.
\item It would be interesting to see how the BPS flow equations derived in sec.~\ref{hyper} arise
precisely in the general classification scheme of supersymmetric solutions obtained in \cite{Bellorin:2007yp}.
\item We have checked that our central charge \eqref{c-final} agrees with the results
of \cite{Benini:2012cz,Benini:2013cda}, where black string solutions corresponding to D3-branes at a
Calabi-Yau singularity have been studied in detail. It may be of some interest to use the flow equations
obtained in section \ref{hyper} to study more complicated type IIB configurations,
as was initiated in \cite{Amariti:2016mnz}.
\end{itemize}

Work along these directions is in progress.

\section*{Acknowledgements}

This work was supported partly by INFN.

\appendix

\section*{\label{Appendix}Appendix}

\section{Useful relations in very special geometry}
\label{app-very-special}

We list here some useful relations that can be proven using the techniques of very special
geometry:
\begin{equation}
\partial_i h_I = -\frac23 G_{IJ}\partial_i h^J\,, \qquad h_I = \frac23 G_{IJ}h^J\,, \qquad
G_{IJ} = \frac92 h_I h_J - \frac12 C_{IJK}h^K\,, \label{eq:veryspecial}
\end{equation}
\begin{equation}
\begin{split}
\mathcal G^{ij}\partial_i h^I\partial_j h^J = G^{IJ} & -\frac23 h^I h^J\,, \qquad\mathcal G^{ij}\partial_i h_I
\partial_j h_J = \frac49 G_{IJ} - \frac23 h_I h_J\,, \\
&\mathcal G^{ij}\partial_i h^I\partial_j h_J = -\frac23\delta^I_J + \frac23 h^I h_J\,.
\label{eq:veryspecial1}
\end{split}
\end{equation}
In the special case where the tensor $T_{ijk}$ that determines the Riemann tensor of the vector multiplet
scalar manifold $\cal M$ (cf.~\cite{Gunaydin:1983bi} for details) is covariantly constant\footnote{This
implies that $\cal M$ is a locally symmetric space.}, one has also
\begin{equation}
C_{IJK} C_{J'\left(LM\right.} C_{\left.PQ\right)K'}\delta^{JJ'}\delta^{KK'} = \frac43\delta_{I\left(L\right.}
C_{\left.MPQ\right)}\,, \label{adjoint-id}
\end{equation}
which is the adjoint identity of the associated Jordan algebra \cite{Gunaydin:1983bi}. Using
\eqref{adjoint-id} and defining $C^{IJK}\equiv\delta^{II'}\delta^{JJ'}\delta^{KK'}C_{I'J'K'}$, one can show
that
\begin{equation}
G^{IJ} = -6 C^{IJK} h_K + 2h^I h^J\,.
\end{equation}

\section{Down to $d= 4$ via $r$-map}
\label{down}

A natural question arising in the discussion of \eqref{eq:hyperflow1} is the relation with the flow
equations of \cite{Halmagyi:2013sla,Klemm:2016wng}, coming from the abelian gauged supergravity
theory in $d=4$. An interesting way to answer this question is to extend the general $r$-map
construction in ungauged supergravity \cite{Cortes:2011aj} to the gauged case.

\subsection{Construction of the $r$-map}

The first step is a Kaluza-Klein reduction along the $z$-direction (i.e., along the string), by using the
ansatz\footnote{In this subsection $\mu,\nu,\ldots$ are curved indices for the four-dimensional theory,
and the dilaton is related to the function $T$ in \eqref{eq:ans} by $T=-\phi/\sqrt3$. Further details on
the notation and the theory in $d=4$ can be found in \cite{Klemm:2016wng}.} 
\begin{equation}
\D s^2_5 = e^{\frac{\phi}{\sqrt3}}\D s^2_4 + e^{-\frac2{\sqrt3}\phi}(\D z + K_\mu\D x^\mu)^2\,,
\qquad A^I = B^I\D z + C^I_\mu\D x^\mu + B^I K_\mu\D x^\mu\,. \label{KK5to4}
\end{equation}
Defining $K_{\mu\nu}=\partial_\mu K_\nu-\partial_\nu K_\mu$ and $C^I_{\mu\nu}=\partial_\mu
C^I_\nu-\partial_\nu C^I_\mu$, the five-dimensional Lagrangian \eqref{eq:genlaghyper} reduces
to\footnote{We choose $\epsilon^{\mu\nu\rho\sigma z}_5=-\epsilon^{\mu\nu\rho\sigma}_4$.}
\begin{equation}
\begin{split}
e_4^{-1}\mathscr L^{(4)} = \,& \frac{R^{(4)}}2 - \frac18 e^{-\sqrt3\phi} K^{\mu\nu} K_{\mu\nu} - \frac14
G_{IJ} e^{-\frac{\phi}{\sqrt3}}(C^{I\mu\nu} + B^I K^{\mu\nu})(C^J_{\mu\nu} + B^J K_{\mu\nu})\\
& -\frac12 e^{\frac{2\phi}{\sqrt3}} G_{IJ}\partial_{\mu} B^I\partial^{\mu} B^J - \frac12 G_{IJ}\partial_{\mu}
h^I\partial^{\mu} h^J - \frac14\partial_{\mu}\phi\partial^{\mu}\phi - h_{uv}\hat{\partial}_{\mu} q^u \hat{\partial}^{\mu} q^v\\
& -\frac{e_4^{-1}}{16}\epsilon^{\mu\nu\rho\sigma} C_{IJK}\big(C^I_{\mu\nu} C^J_{\rho\sigma} B^K +
\frac13 K_{\mu\nu} K_{\rho\sigma} B^I B^J B^K + C^I_{\mu\nu} K_{\rho\sigma} B^J B^K\big)\\
& -e^{\sqrt3\phi}\mathrm{g}^2 B^I k_I^u B^J k_J^v h_{uv} - \mathrm{g}^2 e^{\frac{\phi}{\sqrt3} }V_5\,.
\label{eq: lag4d}
\end{split}
\end{equation}
Now we want to rewrite $\mathscr L^{(4)}$ in the language of $N=2$, $d=4$ supergravity, by using
the identifications of the ungauged case \cite{Ceresole:2007rq}. The coordinates of the special
K\"ahler manifold, K\"ahler potential, K\"ahler metric and electromagnetic field strengths are
given in terms of five-dimensional data respectively by
\begin{equation}
\begin{split}
&z^I = -B^I - i e^{-\frac{\phi}{\sqrt3}} h^I\,, \qquad e^{\cal K} = \frac18 e^{\sqrt3\phi}\,, \\
& g_{I\bar J} = \frac12 e^{\frac{2\phi}{\sqrt3}} G_{IJ}\,, \qquad F_{\mu\nu}^{\Lambda} =
\frac1{\sqrt2} (K_{\mu\nu}, C^I_{\mu\nu})\,,
\end{split}
\label{eq: rmapping}
\end{equation}
where capital greek indices $\Lambda,\Sigma,\ldots$ range from $0$ to $n_{\text v}+1$. If we introduce
the matrices
\begin{equation}
R_{\Lambda\Sigma} =
-\begin{pmatrix}
\frac 13 B & \frac 12 B_J \\
\frac 12 B_I & B_{IJ}
\end{pmatrix}\,,
\qquad
I_{\Lambda\Sigma} = -e^{-\sqrt3\phi}
\begin{pmatrix}
1 + 4 g & 4 g_ {\bar J} \\
4 g_I & 4 g_{I\bar J}
\end{pmatrix}\,,
\end{equation} 
where we defined
\begin{equation}
\begin{split}
B_{IJ} = & C_{IJK} B^K\,, \qquad B_I = C_{IJK} B^J B^K\,, \qquad B = C_{IJK} B^I B^J B^K\,, \\
& g = g_{I\bar J} B^I B^J\,, \qquad g_{I\bar J} B^J = g_I = g_{\bar I} = g_{\bar I J} B^J\,,
\end{split}
\end{equation}
the Lagrangian \eqref{eq: lag4d} can be cast into the form
\begin{equation}
\begin{split}
e_4^{-1}\mathscr L^{(4)} = &\frac R2 - g_{I\bar J}\partial_{\mu} z^I\partial^{\mu}\bar z^{\bar J}
- h_{uv}\hat{\partial}_{\mu} q^u\hat{\partial}^{\mu} q^v \\
& +\frac14 I_{\Lambda\Sigma} F^{\Lambda\mu\nu} F^{\Sigma}_{\mu\nu}
+ \frac18 e_4^{-1}\epsilon^{\mu\nu\rho\sigma} R_{\Lambda\Sigma} F^{\Lambda}_{\mu\nu}
F^{\Sigma}_{\rho\sigma} - \tilde{V}\,,
\end{split}
\end{equation}
with the four-dimensional potential given by
\begin{equation}
\tilde{V} = \mathrm{g}^2 e^{\frac{\phi}{\sqrt3}} V_5 + e^{\sqrt3\phi}\mathrm{g}^2 h_{uv} k_I^u k_J^v
B^I B^J\,. \label{eq:potn}
\end{equation}
The underlying prepotential of the special K\"ahler manifold turns out to be
\begin{equation}
F = \frac16\frac{C_{IJK} X^I X^J X^K}{X^0}\,, \label{eq:prepo}
\end{equation}
chosen the parametrization $X^I/X^0=z^I=-B^I-ie^{-\phi/\sqrt3}h^I$ \cite{Ceresole:2007rq}.

The actual novelties with respect to the ungauged case are the potential and the covariant derivative
acting on the hyperscalars. The former reads
\begin{eqnarray}
\frac{\tilde{V}}{\mathrm{g}^2} &=& -9 e^{\frac{\phi}{\sqrt3}} P^x_I P^x_J\left(h^I h^J - \frac12 G^{IJ}\right) 
+ 9 e^{\frac{\phi}{\sqrt3}} h_{uv} k^u_I k^v_J h^I h^J + 9 e^{\sqrt3\phi} h_{uv} k_I^u k_J^v B^I B^J
\nonumber \\
&=& 18 P^x_I P^x_J\left(\frac14 e^{\frac{\phi}{\sqrt3}} G^{IJ} + \frac12 e^{\sqrt3\phi} B^I B^J
- 4\frac{e^{\sqrt3\phi}}8 (e^{-\frac{\phi}{\sqrt3}} h^I)(e^{-\frac{\phi}{\sqrt3}} h^J)
- \frac12 e^{\sqrt3\phi} B^I B^J\right) \nonumber \\
&& + 72\frac{e^{\sqrt3\phi}}8 h_{uv} k_I^u k_J^v (e^{-\frac{2\phi}{\sqrt3}} h^I h^J + B^I B^J)\,.
\label{eq:potrmap} 
\end{eqnarray}
Now the first two terms in the second line of \eqref{eq:potrmap} combine to give
$-\frac12 I^{\Lambda\Sigma}$ (the inverse of $I_{\Lambda\Sigma}$ defined above), while the last two
terms yield $-4X^I\bar X^J$. Fixing furthermore $\mathrm{g}_4=3\sqrt2\mathrm{g}$, one has thus
\begin{equation}
\begin{split}
\tilde{V} = &\mathrm{g}_4^2\left[P^x_\Lambda P^x_\Sigma\left(-\frac12 I^{\Lambda \Sigma} -
4 X^{\Lambda}\bar{X}^{\Sigma}\right) + 4 h_{uv} k_\Lambda^u k_\Sigma^v X^\Lambda\bar{X}^\Sigma
\right]\Big|_{P^x_0 = 0, k^u_0 = 0} \\
= &\mathrm{g}_4^2 V_4\Big|_{P^x_0 = 0, k^u_0 = 0}\,,
\label{eq:trunpot}
\end{split}
\end{equation}
which is precisely the truncated potential of the four-dimensional theory.\\
The final point to take care of is the covariant derivative of the hyperscalars,
\begin{equation}
\hat{\partial}_{\mu} q^u = \partial_{\mu} q^u + 3\mathrm{g} C^I_\mu k^u_I = \partial_{\mu} q^u + 
\mathrm{g}_4 A^I_{4\mu} k^u_I\,.
\end{equation}
We have therefore shown that the $r$-map can be extended to the case of gauged supergravity, where
the scalar fields have a potential.

\subsection{Comparison with the flow in $N=2$, $d=4$ gauged supergravity}
\label{flows}

The result of the preceeding subsection is completely general and interesting by itself, however our aim is
to use this mapping to compare the flow \eqref{eq:hyperflow} for a black string in $d= 5$ with the
flow equations for black holes in four dimensions obtained in \cite{Halmagyi:2013sla,Klemm:2016wng}.
The latter are driven by the Hamilton-Jacobi function
\begin{equation}
W_4 = e^U\mathrm{Re}(e^{-i\alpha}\mathcal Z_4) - \kappa\mathrm{g}_4 e^{2\psi - U} \mathrm{Im}
(e^{-i\alpha}\mathcal L_4)\,, \label{eq:Wnd}
\end{equation}
where the phase $\alpha$ is defined by
\begin{equation}
e^{2i\alpha} = \frac{\mathcal Z_4 + i\kappa\mathrm{g}_4 e^{2(\psi - U)}\mathcal L_4}{\bar{\mathcal Z}_4
- i\kappa\mathrm{g}_4 e^{2(\psi - U)}\bar{\mathcal L}_4}\,.
\label{eq:alpha}
\end{equation}
Specifying to a purely magnetic charge configuration $\hat{p}^\Lambda=(0,p^I/\sqrt2)$, purely electric 
couplings with $P^x_0=0,k^u_0=0$, and restricting to imaginary scalars, $z^I=-ie^{-\phi/\sqrt3}h^I$,
the quantities defining \eqref{eq:Wnd} become
\begin{equation}
\begin{split}
&\mathcal Z_4 = \frac{3}{2\sqrt2} e^{T/2}\hat{p}^I h_I = \frac34 e^{T/2}\ma Z\,, \qquad
\ma Q^x_4 = P^x_I\hat{p}^I = \frac1{\sqrt2}\ma Q^x\,, \\
&\ma W^x_4 = -\frac i{2\sqrt2} e^{-T/2} P^x_I h^I = -\frac i{2\sqrt2} e^{-T/2}\ma W^x\,, \qquad
\ma L_4 = \ma Q^x_4\ma W^x_4\,,
\end{split}
\end{equation}
where the quantities $\ma Z$, $\ma Q^x$ and $\ma W^x$ were defined in section \ref{hyper}.
Note that axions are absent, since for magnetically
charged black strings the $z$-components $B^I$ of the five-dimensional gauge potentials vanish.
For this choice, \eqref{eq:alpha} becomes $e^{2 i\alpha}=1$. Moreover, taking in account that
$\mathrm{g}_4=3\sqrt2\mathrm{g}$ and choosing $e^{i\alpha}=-1$, the function \eqref{eq:Wnd} boils 
down to \eqref{eq:mainHJhyper}. On the other hand, the Hamilton-Jacobi equation satisfied by
\eqref{eq:Wnd}, namely (3.40) of \cite{Klemm:2016wng}, becomes \eqref{eq:hjbshyper} once the dictionary
is imposed. This proves the expected equivalence between the flows in five and four dimensions.


\begin{thebibliography}{99}

\bibitem{Andrianopoli:2012ee}
  L.~Andrianopoli, R.~D'Auria, P.~Giaccone and M.~Trigiante,
  ``Rotating black holes, global symmetry and first order formalism,''
  JHEP {\bf 1212} (2012) 078
  [arXiv:1210.4047 [hep-th]].

\bibitem{Hristov:2012nu}
  K.~Hristov, S.~Katmadas and V.~Pozzoli,
  ``Ungauging black holes and hidden supercharges,''
  JHEP {\bf 1301} (2013) 110
  [arXiv:1211.0035 [hep-th]].

\bibitem{Andrianopoli:2009je}
  L.~Andrianopoli, R.~D'Auria, E.~Orazi and M.~Trigiante,
  ``First order description of $d=4$ static black holes and the Hamilton-Jacobi equation,''
  Nucl.\ Phys.\ B {\bf 833} (2010) 1
  [arXiv:0905.3938 [hep-th]].

\bibitem{Trigiante:2012eb}
  M.~Trigiante, T.~Van Riet and B.~Vercnocke,
  ``Fake supersymmetry versus Hamilton-Jacobi,''
  JHEP {\bf 1205} (2012) 078
  [arXiv:1203.3194 [hep-th]].

\bibitem{Miller:2006ay}
  C.~M.~Miller, K.~Schalm and E.~J.~Weinberg,
  ``Nonextremal black holes are BPS,''
  Phys.\ Rev.\ D {\bf 76} (2007) 044001
  [hep-th/0612308].

\bibitem{Ceresole:2007wx}
  A.~Ceresole and G.~Dall'Agata,
  ``Flow equations for non-BPS extremal black holes,''
  JHEP {\bf 0703} (2007) 110
  [hep-th/0702088].

\bibitem{LopesCardoso:2007qid}
  G.~Lopes Cardoso, A.~Ceresole, G.~Dall'Agata, J.~M.~Oberreuter and J.~Perz,
  ``First-order flow equations for extremal black holes in very special geometry,''
  JHEP {\bf 0710} (2007) 063
  [arXiv:0706.3373 [hep-th]].

\bibitem{Cardoso:2008gm}
  G.~L.~Cardoso and V.~Grass,
  ``On five-dimensional non-extremal charged black holes and FRW cosmology,''
  Nucl.\ Phys.\ B {\bf 803} (2008) 209
  [arXiv:0803.2819 [hep-th]].

\bibitem{Dall'Agata:2010gj}
  G.~Dall'Agata and A.~Gnecchi,
  ``Flow equations and attractors for black holes in $N=2$ $\text{U}(1)$ gauged supergravity,''
  JHEP {\bf 1103} (2011) 037
  [arXiv:1012.3756 [hep-th]].

\bibitem{Galli:2011fq}
  P.~Galli, T.~Ort\'{\i}n, J.~Perz and C.~S.~Shahbazi,
  ``Non-extremal black holes of $N=2$, $d=4$ supergravity,''
  JHEP {\bf 1107} (2011) 041
  [arXiv:1105.3311 [hep-th]].

\bibitem{Barisch:2011ui}
  S.~Barisch, G.~Lopes Cardoso, M.~Haack, S.~Nampuri and N.~A.~Obers,
  ``Nernst branes in gauged supergravity,''
  JHEP {\bf 1111} (2011) 090
  [arXiv:1108.0296 [hep-th]].

\bibitem{BarischDick:2012gj}
  S.~Barisch-Dick, G.~Lopes Cardoso, M.~Haack and S.~Nampuri,
  ``Extremal black brane solutions in five-dimensional gauged supergravity,''
  JHEP {\bf 1302} (2013) 103
  [arXiv:1211.0832 [hep-th]].

\bibitem{Klemm:2012vm}
  D.~Klemm and O.~Vaughan,
  ``Nonextremal black holes in gauged supergravity and the real formulation of special geometry II,''
  Class.\ Quant.\ Grav.\  {\bf 30} (2013) 065003
  [arXiv:1211.1618 [hep-th]].

\bibitem{Gnecchi:2012kb}
  A.~Gnecchi and C.~Toldo,
  ``On the non-BPS first order flow in $N=2$ $\text{U}(1)$-gauged supergravity,''
  JHEP {\bf 1303} (2013) 088
  [arXiv:1211.1966 [hep-th]].

\bibitem{Gnecchi:2014cqa}
  A.~Gnecchi and C.~Toldo,
  ``First order flow for non-extremal AdS black holes and mass from holographic renormalization,''
  JHEP {\bf 1410} (2014) 075
  [arXiv:1406.0666 [hep-th]].

\bibitem{Cardoso:2015wcf}
  G.~L.~Cardoso, M.~Haack and S.~Nampuri,
  ``Nernst branes with Lifshitz asymptotics in $N=2$ gauged supergravity,''
  JHEP {\bf 1606} (2016) 144
  [arXiv:1511.07676 [hep-th]].

\bibitem{Klemm:2016wng}
  D.~Klemm, N.~Petri and M.~Rabbiosi,
  ``Symplectically invariant flow equations for $N=2$, $D=4$ gauged supergravity with hypermultiplets,''
  JHEP {\bf 1604} (2016) 008
  [arXiv:1602.01334 [hep-th]].

\bibitem{Ceresole:2001wi}
  A.~Ceresole, G.~Dall'Agata, R.~Kallosh and A.~Van Proeyen,
  ``Hypermultiplets, domain walls and supersymmetric attractors,''
  Phys.\ Rev.\ D {\bf 64} (2001) 104006
  [hep-th/0104056].

\bibitem{Maldacena:2000mw}
  J.~M.~Maldacena and C.~Nu\~nez,
  ``Supergravity description of field theories on curved manifolds and a no go theorem,''
  Int.\ J.\ Mod.\ Phys.\ A {\bf 16} (2001) 822
  [hep-th/0007018].

\bibitem{Benini:2012cz}
  F.~Benini and N.~Bobev,
  ``Exact two-dimensional superconformal R-symmetry and c-extremization,''
  Phys.\ Rev.\ Lett.\  {\bf 110} (2013) no.6,  061601
  [arXiv:1211.4030 [hep-th]].

\bibitem{Benini:2015bwz}
  F.~Benini, N.~Bobev and P.~M.~Crichigno,
  ``Two-dimensional SCFTs from D3-branes,''
  JHEP {\bf 1607} (2016) 020
  [arXiv:1511.09462 [hep-th]].

\bibitem{Karndumri:2013iqa}
  P.~Karndumri and E.~O Colgain,
  ``Supergravity dual of $c$-extremization,''
  Phys.\ Rev.\ D {\bf 87} (2013) no.10,  101902
  [arXiv:1302.6532 [hep-th]].

\bibitem{Amariti:2016mnz}
  A.~Amariti and C.~Toldo,
  ``Betti multiplets, flows across dimensions and c-extremization,''
  arXiv:1610.08858 [hep-th].

\bibitem{Gunaydin:1983bi}
  M.~G\"unaydin, G.~Sierra and P.~K.~Townsend,
  ``The geometry of $N=2$ Maxwell-Einstein supergravity and Jordan algebras,''
  Nucl.\ Phys.\ B {\bf 242} (1984) 244.

\bibitem{Gunaydin:1984ak}
  M.~G\"unaydin, G.~Sierra and P.~K.~Townsend,
  ``Gauging the $d=5$ Maxwell-Einstein supergravity theories: More on Jordan algebras,''
  Nucl.\ Phys.\ B {\bf 253} (1985) 573.

\bibitem{deWit:1992cr}
  B.~de Wit and A.~Van Proeyen,
  ``Broken sigma model isometries in very special geometry,''
  Phys.\ Lett.\ B {\bf 293} (1992) 94
  [hep-th/9207091].

\bibitem{Cadavid:1995bk}
  A.~C.~Cadavid, A.~Ceresole, R.~D'Auria and S.~Ferrara,
  ``Eleven-dimensional supergravity compactified on Calabi-Yau threefolds,''
  Phys.\ Lett.\ B {\bf 357} (1995) 76
  [hep-th/9506144].

\bibitem{Berkooz:2008rj}
  M.~Berkooz and B.~Pioline,
  ``5D black holes and non-linear sigma models,''
  JHEP {\bf 0805} (2008) 045
  [arXiv:0802.1659 [hep-th]].

\bibitem{Cortes:2011aj}
  V.~Cort\'es, T.~Mohaupt and H.~Xu,
  ``Completeness in supergravity constructions,''
  Commun.\ Math.\ Phys.\  {\bf 311} (2012) 191
  [arXiv:1101.5103 [hep-th]].

\bibitem{Cacciatori:2003kv}
  S.~L.~Cacciatori, D.~Klemm and W.~A.~Sabra,
  ``Supersymmetric domain walls and strings in $D=5$ gauged supergravity coupled to vector multiplets,''
  JHEP {\bf 0303} (2003) 023
  [hep-th/0302218].

\bibitem{Klemm:2000nj}
  D.~Klemm and W.~A.~Sabra,
  ``Supersymmetry of black strings in $D=5$ gauged supergravities,''
  Phys.\ Rev.\ D {\bf 62} (2000) 024003
  [hep-th/0001131].

\bibitem{Ortin:1996bz}
  T.~Ort\'{\i}n,
  ``Extremality versus supersymmetry in stringy black holes,''
  Phys.\ Lett.\ B {\bf 422} (1998) 93
  [hep-th/9612142].

\bibitem{Halmagyi:2013qoa}
  N.~Halmagyi,
  ``BPS black hole horizons in $N=2$ gauged supergravity,''
  JHEP {\bf 1402} (2014) 051
  [arXiv:1308.1439 [hep-th]].

\bibitem{Chakrabarty:2016mpf}
  B.~Chakrabarty, K.~Inbasekar and R.~Samanta,
  ``On the supersymmetry of Bianchi attractors in gauged supergravity,''
  arXiv:1610.03033 [hep-th].

\bibitem{Benini:2013cda}
  F.~Benini and N.~Bobev,
  ``Two-dimensional SCFTs from wrapped branes and c-extremization,''
  JHEP {\bf 1306} (2013) 005
  [arXiv:1302.4451 [hep-th]].

\bibitem{Hristov:2014eza}
  K.~Hristov,
  ``Dimensional reduction of BPS attractors in AdS gauged supergravities,''
  JHEP {\bf 1412} (2014) 066
  [arXiv:1409.8504 [hep-th]].

\bibitem{Brown:1986nw}
  J.~D.~Brown and M.~Henneaux,
  ``Central charges in the canonical realization of asymptotic symmetries: An example from
  three-dimensional gravity,''
  Commun.\ Math.\ Phys.\  {\bf 104} (1986) 207.

\bibitem{Chimento:2015rra}
  S.~Chimento, D.~Klemm and N.~Petri,
  ``Supersymmetric black holes and attractors in gauged supergravity with hypermultiplets,''
  JHEP {\bf 1506} (2015) 150
  [arXiv:1503.09055 [hep-th]].

\bibitem{Bernamonti:2007bu}
  A.~Bernamonti, M.~M.~Caldarelli, D.~Klemm, R.~Olea, C.~Sieg and E.~Zorzan,
  ``Black strings in AdS$_5$,''
  JHEP {\bf 0801} (2008) 061
  [arXiv:0708.2402 [hep-th]].

\bibitem{Bellorin:2007yp}
  J.~Bellor\'{\i}n and T.~Ort\'{\i}n,
  ``Characterization of all the supersymmetric solutions of gauged $N=1$, $d=5$ supergravity,''
  JHEP {\bf 0708} (2007) 096
  [arXiv:0705.2567 [hep-th]].

\bibitem{Halmagyi:2013sla}
  N.~Halmagyi, M.~Petrini and A.~Zaffaroni,
  ``BPS black holes in AdS$_4$ from M-theory,''
  JHEP {\bf 1308} (2013) 124
  [arXiv:1305.0730 [hep-th]].

\bibitem{Ceresole:2007rq}
  A.~Ceresole, S.~Ferrara and A.~Marrani,
  ``4d/5d correspondence for the black hole potential and its critical points,''
  Class.\ Quant.\ Grav.\  {\bf 24} (2007) 5651
  [arXiv:0707.0964 [hep-th]].






\end{thebibliography}
\end{document}